\documentclass[twocolumn,aps,showpacs,preprintnumbers,amsmath,amssymb]{revtex4-1}

\usepackage{bm}
\usepackage{tabularx}
\usepackage{graphicx}
\usepackage{color}

\begin{document}

\title{Heat transport in RbFe$_2$As$_2$ single crystal: evidence for nodal superconducting gap}

\author{Z. Zhang,$^1$ A. F. Wang,$^2$ X. C. Hong,$^1$ J. Zhang,$^1$ B. Y. Pan,$^1$ J. Pan,$^1$ Y. Xu,$^1$ X. G. Luo,$^{2,3}$ X. H. Chen,$^{2,3}$ and S. Y. Li$^{1,3,*}$}

\affiliation{$^1$State Key Laboratory of Surface Physics, Department
of Physics, and Laboratory of Advanced Materials, Fudan University,
Shanghai 200433, P. R. China\\
$^2$Hefei National Laboratory for Physical Science at Microscale and
Department of Physics, University of Science and Technology of
China, Hefei, Anhui 230026, P. R. China\\
$^3$Collaborative Innovation Center of Advanced Microstructures, Nanjing 210093, P. R. China}

\date{\today}

\begin{abstract}

The in-plane thermal conductivity of iron-based superconductor RbFe$_2$As$_2$ single crystal ($T_c \approx$ 2.1 K) was measured down to 100 mK.
In zero field, the observation of a significant residual linear term $\kappa_0/T$ = 0.65 mW K$^{-2}$ cm$^{-1}$ provides clear evidence for nodal superdonducting gap. The field dependence of $\kappa_0/T$ is similar to that of its sister compound CsFe$_2$As$_2$ with comparable residual resistivity $\rho_0$, and lies between the dirty and clean KFe$_2$As$_2$. These results suggest that the (K,Rb,Cs)Fe$_2$As$_2$ serial superconductors have a common nodal gap structure.

\end{abstract}

\pacs{74.70.Xa, 74.25.fc}

\maketitle

\section{Introduction}
The iron-based superconductors \cite{Kamihara,XHChen} have attracted great attention since Hosono and co-workers reported the discovery of 26 K superconductivity in fluorine doped LaFeAsO in 2008 \cite{Kamihara}. Unfortunately, there is still no consensus on the superconducting mechanism in them, mainly due to their complicated electronic structures \cite{IIMazin,HDing1,HPJ}.

There are many families of the iron-based superconductors, such as LaO$_{1-x}$F$_{x}$FeAs (``1111''), Ba$_{1-x}$K$_{x}$Fe$_{2}$As$_{2}$ (``122''), NaFe$_{1-x}$Co$_x$As (``111''), and FeSe$_{x}$Te$_{1-x}$ (``11'') \cite{Stewart}. Among them,  the ``122'' family is the most studied one due to the easy growth of sizable high-quality single crystals \cite{XFWang}. Intriguingly, the members of this family do not share a universal superconducting gap structure. While the optimally doped Ba$_{0.6}$K$_{0.4}$Fe$_2$As$_2$ and BaFe$_{1.85}$Co$_{0.15}$As$_2$ have nodeless superconducting gaps \cite{HDing1,KTerashima,XGLuo,Tanatar}, the extremely hole-doped KFe$_2$As$_2$ was reported to be a nodal superconductor \cite{JKDong,KHashimoto}. Furthermore, the isovalently doped BaFe$_2$(As$_{1-x}$P$_x$)$_2$ \cite{YNakai,KHashimoto1,YZhang} and Ba(Fe$_{1-x}$Ru$_x$)$_2$As$_2$ \cite{XQiu} also manifest nodal superconductivity. So far, the origin of these nodal superconducting gaps is still under debate, particularly in KFe$_2$As$_2$ \cite{JKDong,KHashimoto,JReid,FTafti,KOkazaki}. The detailed thermal conductivity study provided compelling evidences for a $d$-wave gap in KFe$_2$As$_2$ \cite{JReid}, but the low-temperature angle-resolved photoemission spectroscopy (ARPES) measurements clearly showed nodal $s$-wave gap \cite{KOkazaki}. Recent ARPES and thermal conductivity experiments on highly hole-doped Ba$_{1-x}$K$_{x}$Fe$_{2}$As$_{2}$ also support nodal $s$-wave gap \cite{NXu,XCHongnew}.

KFe$_2$As$_2$ has two sister compounds, CsFe$_2$As$_2$ and RbFe$_2$As$_2$, and both of them are superconducting \cite{Sasmal,ZBukowski}. While muon-spin spectroscopy measurements on RbFe$_2$As$_2$ polycrystals suggested that RbFe$_2$As$_2$ is best described by
a two-gap $s$-wave model \cite{ZShermadini1,ZShermadini2}, recent specific heat and thermal conductivity measurements on CsFe$_2$As$_2$ single crystals provided clear evidences for nodal superconducting gap in CsFe$_2$As$_2$ \cite{AFWang,XCHong}. To clarify whether the superconducting gap structure of RbFe$_2$As$_2$ is indeed different from those of KFe$_2$As$_2$ and CsFe$_2$As$_2$, more experiments on RbFe$_2$As$_2$ single crystals are highly desired.

In this paper, we present the low-temperature thermal conductivity of RbFe$_2$As$_2$ single crystal down to 100 mK. A significant residual linear term $\kappa_0/T$ = 0.65 $\pm$ 0.03 mW K$^{-2}$ cm$^{-1}$ is obtained in zero magnetic field, and the field dependence of $\kappa_0/T$ mimics that of CsFe$_2$As$_2$. These results clarify that RbFe$_2$As$_2$ is also a nodal superconductor. The three compounds KFe$_2$As$_2$, RbFe$_2$As$_2$, and CsFe$_2$As$_2$ may have a common superconducting gap structure.

\section{Experiment}

The RbFe$_2$As$_2$ single crystals were grown by self-flux method for the first time, and the process is the same as the growth of CsFe$_2$As$_2$ single crystals \cite{AFWang}. The dc magnetization was measured using a superconducting quantum interference device (MPMS, Quantum design).
The specific heat measurement above 1.9 K was performed in a physical property measurement system (PPMS, Quantum design) via the relaxation method,
and below 1.9 K it was measured in a small dilution refrigerator integrated into the PPMS.
For transport measurements, the RbFe$_2$As$_2$ single crystal was cleaved to a rectangular shape of dimensions 2.2 $\times$ 1.0 mm$^2$ in the $ab$ plane, with 40 $\mu$m thickness along the $c$ axis. Contacts were made directly on the sample surfaces with silver paint (Dupont 4929N), which were used for both resistivity and thermal conductivity measurements. To avoid degradation, the sample was exposed in air for less than 2 hours. The contacts are metallic with a typical resistance 100 m$\Omega$ at 2 K. In-plane thermal conductivity was measured in a dilution refrigerator, using a standard four-wire steady-state method with two RuO$_2$ chip thermometers, calibrated {\it in situ} against a reference RuO$_2$ thermometer. Magnetic fields were applied along the $c$ axis and perpendicular to the heat current. To ensure a homogeneous field distribution in the sample, all fields were applied at a temperature above $T_c$ for transport measurements.

\section{Results and Discussion}

\begin{figure}
\includegraphics[clip,width=8cm]{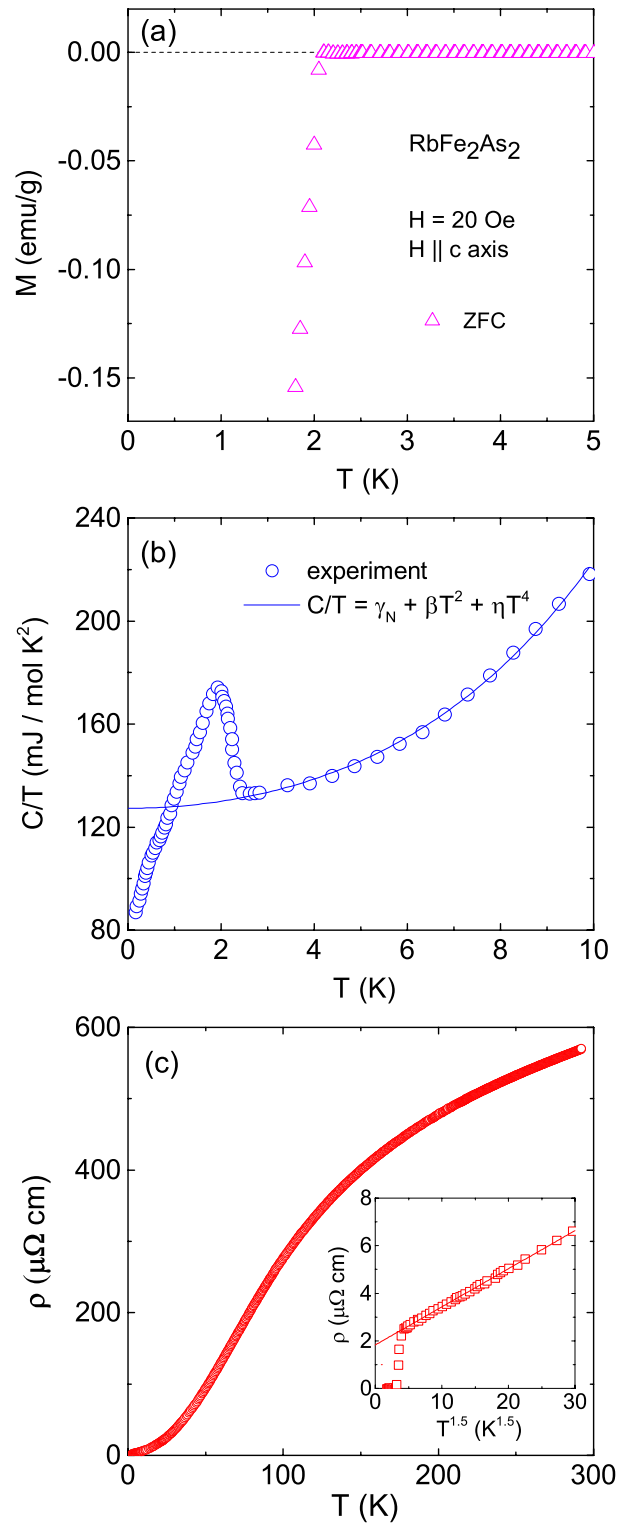}
\caption{(Color online). (a) Low-temperature dc magnetization of RbFe$_2$As$_2$ single crystal in $H =$ 20 Oe along $c$ axis,
with zero-field cooling process.
(b) Temperature dependence of specific heat $C/T$ for RbFe$_2$As$_2$ single crystal in zero field, plotted as $C/T$ vs $T$.
The solid line is the best fit to $C_{normal}$ = $\gamma_NT$ + $\beta$$T^{3}$+ $\eta$$T^{5}$ from 2.4 to 10 K.
(c) In-plane resistivity of RbFe$_2$As$_2$ single crystal in zero field. The data between 2.2 and 9 K can be fitted to $\rho$ = $\rho_0$ + $AT^{1.5}$,
as shown in the inset, which gives $\rho_{0}$ = 1.84 $\mu$$\Omega$ cm.}
\end{figure}

\begin{figure}
\includegraphics[clip,width=8cm]{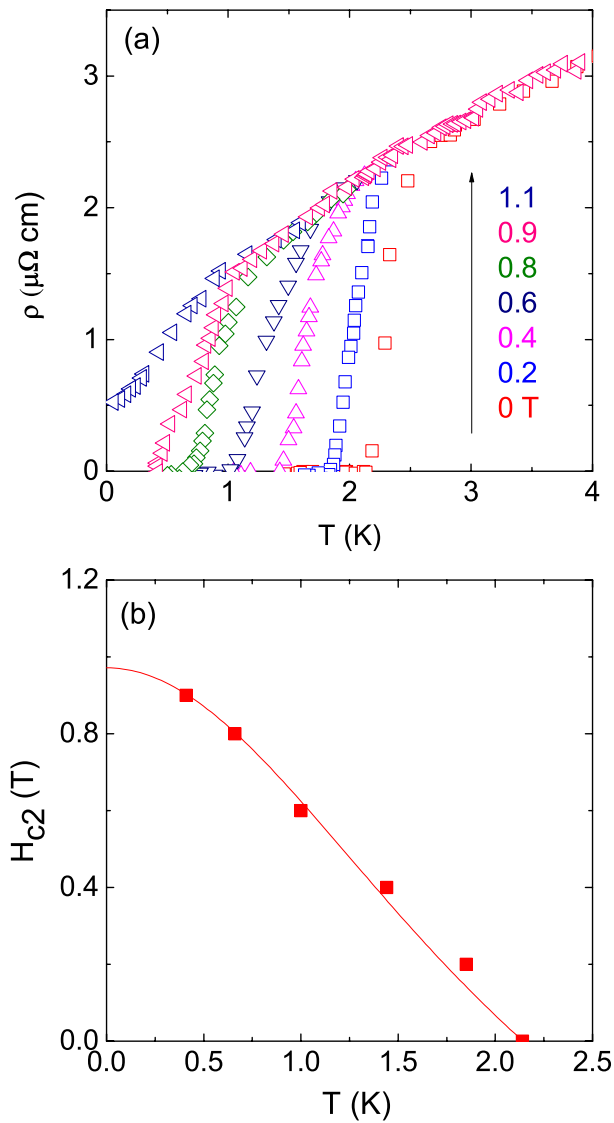}
\caption{(Color online). (a) Low-temperature resistivity of
RbFe$_2$As$_2$ single crystal in magnetic fields up to 1.1 T. (b) Temperature dependence of the upper critical field $H_{c2}(T)$,
defined by $\rho = 0$ in (a). The solid line is a fit of $H_{c2}(T)$ to the Ginzburg-Landau equation, which gives
 $H_{c2}(0) \approx$ 0.97 T.}
\end{figure}

Figure 1(a) shows the low-temperature dc magnetization of RbFe$_2$As$_2$ single crystal, measured in $H$ = 20 Oe along $c$ axis,
with zero-field cooling process. The $T_c \approx$ 2.10 K is defined at the onset of the diamagnetic transition. The magnetization does not saturate down to 1.8 K, where the superconducting volume fraction is already as large as 40\%. With decreasing temperature, the superconducting volume fraction should further increase to reach the fully shielded state.

In Fig. 1(b), we present the low-temperature specific heat of RbFe$_2$As$_2$ single crystal down to 100 mK in zero field, plotted as $C/T$ vs $T$.
A significant jump due to the superconducting transition is observed at $T_c \approx$ 2.1 K, which indicates the high quality of our sample.
In order to determine the zero-field normal-state Sommerfeld coefficient $\gamma_N$, the specific heat above $T_c$ is fitted to $C_{normal}$ = $\gamma_NT$ + $\beta$$T^{3}$+ $\eta$$T^{5}$, with $\beta$ and $\eta$ as the lattice coefficients. The solid line in Fig. 1(b) is the best fit of $C/T$ from 2.4 to 10 K, which gives $\gamma_N$ = 127.3 $\pm$ 0.9 mJ  mol$^{-1}$ K$^{-2}$, $\beta$ = 0.66 $\pm$ 0.04 mJ mol$^{-1}$ K$^{-4}$, and $\eta$ = 0.0029 $\pm$ 0.0005 mJ mol$^{-1}$ K$^{-6}$. From the relation $\theta_D$ = (12$\pi^{4}RZ$ / 5$\beta$)$^{1/3}$, where $R$ is the molar gas constant and $Z$ = 5 is the total number of atoms in one unit cell, the Debye temperature $\theta_D$ = 245 K is estimated. This value is comparable to those of KFe$_2$As$_2$ and CsFe$_2$As$_2$ \cite{Sergey,AFWang}.

The in-plane resistivity of RbFe$_2$A$_2$ single crystal in zero filed is plotted in Fig. 1(c).
The $T_c \approx$ 2.13 K, defined by $\rho$ = 0, agrees well with the magnetization and specific heat measurements.
For the polycrystalline sample of RbFe$_2$A$_2$, $T_c =$ 2.6 K was defined at the onset of the diamagnetic transition \cite{ZBukowski},
which is 0.5 K higher than our single crystal. Similarly, $T_c =$ 2.2 K was defined at the onset of the diamagnetic transition for the CsFe$_2$As$_2$ polycrystal \cite{Sasmal}, but $T_c =$ 1.8 K was found in the CsFe$_2$As$_2$ single crystal \cite{AFWang}. It is unclear why the $T_c$ shows difference between polycrystalline sample and single crystal for RbFe$_2$A$_2$ and CsFe$_2$A$_2$. In case that the single crystals have intrinsic $T_c$, the $T_c$ of (K, Rb, Cs)Fe$_2$A$_2$ series (3.8, 2.1, and 1.8 K, respectively) decreases with the increase of the ionic radius of alkali metal. In the inset of Fig. 1(c), the normal-state $\rho(T)$ below 9 K can be well fitted by $\rho$ = $\rho_0$ + $AT^{1.5}$, with $\rho_0$ = 1.84 $\pm$ 0.01 $\mu$$\Omega$ cm and $A$ = 0.16 $\mu$$\Omega$ cm K$^{-1.5}$. Similar non-Fermi-liquid behavior of $\rho(T)$ was also observed in KFe$_2$As$_2$ and CsFe$_2$As$_2$ \cite{JKDong,JReid,XCHong},
which may result from antiferromagnetic spin fluctuations \cite{SWZhang}. The residual resistivity ratio RRR =  $\rho$(292K)/$\rho_0 \approx$ 310 again reflects the high quality of our RbFe$_2$As$_2$ single crystal.

Figure 2(a) shows the low-temperature resistivity of RbFe$_2$As$_2$ single crystal in magnetic fields up to 1.1 T. In order to estimate the zero-temperature upper critical field $H_{c2}(0)$, the temperature dependence of $H_{c2}(T)$ is plotted in Fig. 2(b), defined by $\rho$ = 0 in Fig. 2(a). $H_{c2}(0) \approx$ 0.97 T is obtained by fitting the data with the Ginzburg-Landau equation $H_{c2}(T) = H_{c2}(0)[1-(T/T_c)^{2}]/[1+(T/T_c)^{2}]$ \cite{JAWoollam,CKJones}.

\begin{figure}
\includegraphics[clip,width=8.4cm]{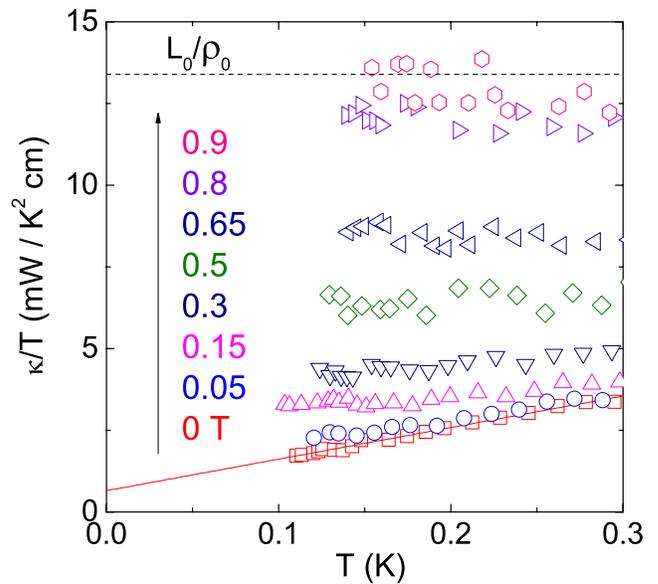}
\caption{(Color online). Low-temperature in-plane thermal
conductivity of RbFe$_2$As$_2$ single crystal in zero and magnetic
fields applied along the $c$ axis. The solid line is a fit of the
zero-field data between 0.1 and 0.3 K to $\kappa/T = a + bT$, giving a residual
linear term $\kappa_0/T$ = 0.65 $\pm$ 0.03 mW K$^{-2}$ cm$^{-1}$. The dashed
line is the normal-state Wiedemann-Franz law expectation
$L_0$/$\rho_0$, with $L_0$ the Lorenz number 2.45 $\times$ 10$^{-8}$
W $\Omega$ K$^{-2}$ and $\rho_0$ = 1.84 $\mu$$\Omega$ cm.}
\end{figure}

The low-temperature heat transport measurement is a bulk technique to probe the gap structure of superconductors \cite{Shakeripour}.
In Fig. 3, the in-plane thermal conductivity of RbFe$_2$As$_2$ single crystal in zero and applied field is plotted as $\kappa/T$ vs $T$ \cite{MSutherland,SYLi}.
The thermal conductivity at very low temperature can be usually fitted to $\kappa/T$ = $a + bT^{\alpha-1}$, in which the two terms  $aT$ and $bT^{\alpha}$ represent contributions from electrons and phonons, respectively. The power $\alpha$ is typically between 2 and 3, due to specular reflections of phonons at the boundary \cite{MSutherland,SYLi}. Since all the curves in Fig. 3 are roughly linear, we fix $\alpha$ to 2. The value $\alpha \approx$ 2 has been previously observed in dirty KFe$_2$As$_2$ \cite{JKDong}, Ba(Fe$_{1-x}$Ru$_x$)$_2$As$_2$ \cite{XQiu}, and CsFe$_2$As$_2$ single crystals \cite{XCHong}. Here, we only focus on the electronic term.

In zero field, the fitting of the data between 0.1 to 0.3 K gives $\kappa_0/T$ = 0.65 $\pm$ 0.03 mW K$^{-2}$cm$^{-1}$.
If we slightly change the fitting range, we obtain $\kappa_0/T$ = 0.62 $\pm$ 0.03 mW K$^{-2}$cm$^{-1}$ for the range below 0.27 K and $\kappa_0/T$ = 0.63 $\pm$ 0.04 mW K$^{-2}$cm$^{-1}$ for the range below 0.24 K. Therefore, the value of $\kappa_0/T$ basically does not depend on the temperature range chosen for the fit. Such a significant $\kappa_0/T$  is usually contributed by nodal quasiparticles, thus it is a strong evidence for nodal superconducting gap \cite{Shakeripour}. Previously, $\kappa_0/T$ = 2.27 $\pm$ 0.02 and 3.6 $\pm$ 0.5 mW K$^{-2}$cm$^{-1}$ were observed for dirty and clean KFe$_2$As$_2$ single crystals, respectively \cite{JReid,JKDong}. For CsFe$_2$As$_2$ single crystal with  $\rho_0$ = 1.80 $\mu$$\Omega$ cm, $\kappa_0/T$ = 1.27 $\pm$ 0.04 mW K$^{-2}$cm$^{-1}$ was found \cite{XCHong}. The zero-field value of $\kappa_0/T$ for RbFe$_2$As$_2$ is about 5$\%$ of the normal-state Widemann-Franz law expectation $\kappa_{N0}/T$ = $L_0/\rho_0$ = 13.5 mW K$^{-2}$ cm$^{-1}$, with $L_0$ the Lorenz number 2.45 $\times$ 10$^{-8}$ W $\Omega$ K$^{-2}$ and $\rho_0$ = 1.84 $\mu$$\Omega$ cm. In $H$ = 0.9 T, the experimental data roughly satisfy the Widemann-Franz law, so we take 0.9 T as the bulk $H_{c2}(0)$.
This value is slightly lower than that obtained from resistivity measurements, but it does not affect our discussion of the filed dependence of $\kappa_0/T$ below.

\begin{figure}
\includegraphics[clip,width=8.7cm]{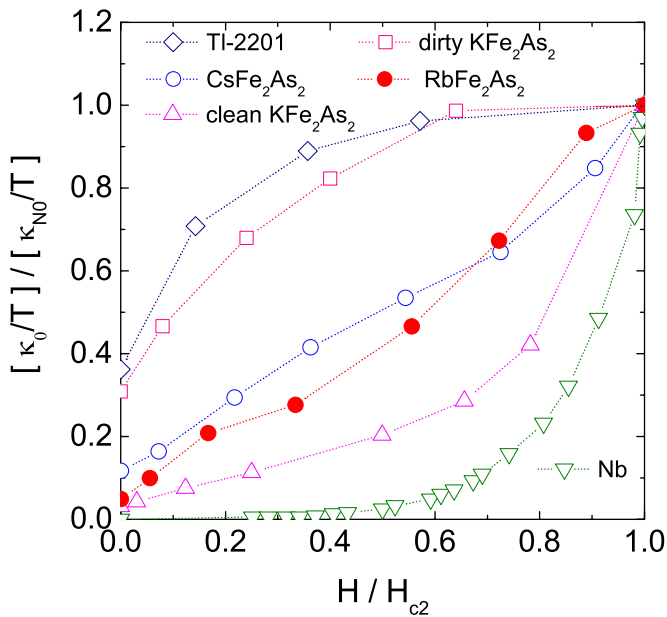}
\caption{(Color online). Normalized residual linear term
$\kappa_0/T$ of RbFe$_2$As$_2$ as a function of $H/H_{c2}$. For
comparison, similar data are shown for the clean $s$-wave
superconductor Nb \cite{Lowell}, the $d$-wave cuprate
superconductor Tl-2201 \cite{Proust}, the dirty and clean KFe$_2$As$_2$ \cite{JKDong,JReid}, and CsFe$_2$As$_2$ \cite{XCHong}.}
\end{figure}

The field dependence of $\kappa_0/T$ may provide more information on the superconducting gap structure \cite{Shakeripour}.
In Fig. 4, we plot the normalized $\kappa_0(H)/T$ of  RbFe$_2$As$_2$ together with the typical $s$-wave superconductor Nb \cite{Lowell},
the $d$-wave cuprate superconductor Tl$_2$Ba$_2$CuO$_{6+\delta}$ (Tl-2201) \cite{Proust}, the dirty and clean KFe$_2$As$_2$ \cite{JKDong,JReid}, and CsFe$_{2}$As$_{2}$ \cite{XCHong}. For an $s$-wave superconductor with isotropic gap, such as Nb, $\kappa_{0}/T$ grows exponentially with the field \cite{Lowell}. For the $d$-wave superconductor Tl-2201, $\kappa_0/T$ increases roughly proportional to $H^{1/2}$ at low field \cite{Proust},
due to the Volovik effect \cite{Volovik}. From Fig. 4, the normalized $\kappa_0(H)/T$ curve of  RbFe$_2$As$_2$ is very close to that of CsFe$_2$As$_2$
and lies between the dirty and clean KFe$_2$As$_2$.

\begin{table}
\centering \caption{The superconducting transiton temperature $T_c$, residual resistivity $\rho_0$, zero-field normal-state Sommerfeld coefficient $\gamma_N$, upper critical field $H_{c2}(0)$, and residual linear term $\kappa_0/T$ of the clean and dirty KFe$_2$As$_2$, CsFe$_2$As$_2$, and RbFe$_2$As$_2$. These values are taken from Refs. \cite{JKDong,JReid,XCHong,AFWang,MAH,JSK,Terashima} and this work.}\label{1}
\begin{tabularx}{0.48\textwidth}{p{2.3cm}p{0.8cm}p{1.3cm}p{1.2cm}p{1.1cm}p{0.9cm}}\hline\hline
                        & ~$T_c$   & ~~~~$\rho_0$  & ~~$\gamma_N$  &$H_{c2}(0)$  &~$\kappa_0/T$         \\
                         & (K) & ($\mu\Omega$ cm) &($\frac{mJ}{mol K^2}$) &~~(T) &($\frac{mW}{K^2 cm}$)   \\ \hline
KFe$_{2}$As$_{2}$(clean)  & 3.8   & ~~~0.21 &~~~94 &~~1.60    & ~~3.60        \\
KFe$_{2}$As$_{2}$(dirty)  & 2.5  & ~~~3.32 &~~~91  &~~1.25    & ~~2.27        \\
RbFe$_{2}$As$_{2}$        & 2.1  & ~~~1.84 &~~127 &~~0.97     & ~~0.65        \\
CsFe$_{2}$As$_{2}$        & 1.8   &~~~1.80 &~~184 &~~1.40     & ~~1.27         \\

\hline \hline
\end{tabularx}
\end{table}

As listed in Table I, the $\rho_0$ of dirty and clean KFe$_2$As$_2$ differ by 15 times \cite{JKDong,JReid}, while RbFe$_2$As$_2$ and CsFe$_2$As$_2$ have comparable $\rho_0$, with values lying between that of dirty and clean KFe$_2$As$_2$. Therefore, in (K,Rb,Cs)Fe$_2$As$_2$ serial superconductors, the field dependence of $\kappa_0/T$ seems to correlate with the impurity level. Although Reid $et$ $al.$ argued that the $\kappa_0(H)/T$ of clean KFe$_2$As$_2$ is a compelling evidence for $d$-wave gap \cite{JReid}, recent thermal conductivity measurements on highly hole-doped Ba$_{1-x}$K$_x$Fe$_2$As$_2$ single crystals support nodal $s$-wave gap \cite{XCHongnew}. For such a complex nodal $s$-wave gap structure, likely with both nodal and nodeless gaps of different magnitudes,
it is hard to get a theoretical curve of $\kappa_0(H)/T$. One needs to carefully consider the effect of impurity on the behavior of $\kappa_0(H)/T$.
Nevertheless, the evolution of the normalized $\kappa_0(H)/T$ suggests a common nodal gap structure in (K,Rb,Cs)Fe$_2$As$_2$ serial superconductors.

In Table I, we also list the $T_c$, $\gamma_N$, $H_{c2}(0)$, and $\kappa_0/T$ of the (K,Rb,Cs)Fe$_2$As$_2$ serial superconductors \cite{JKDong,JReid,XCHong,AFWang,MAH,JSK,Terashima}. Both $T_c$ and $\gamma_N$ show a systematic change with increasing the ionic radii of alkali metal. The $\gamma_N$ values of RbFe$_2$As$_2$ and CsFe$_2$As$_2$ are very large among all iron-based superconductors, which reflects their abnormally large density of states or effective mass of electrons. This may be explained by recent ARPES measurement on CsFe$_2$As$_2$ single crystals, which suggests that the large separation of FeAs layers along $c$ axis makes the system more two-dimensional and enhances the electronic correlations \cite{ZSun}. Neither the $H_{c2}(0)$ nor $\kappa_0/T$ shows a systematic change. The $H_{c2}(0)$ of CsFe$_2$As$_2$ is abnormally high, which may also relate to its much enhanced two dimensionality and electronic correlations. As for the $\kappa_0/T$, it depends on the very details of the nodal gap, such as the slope of the gap at the node. For the accidental nodes appearing in the complex Fermi surfaces of (K,Rb,Cs)Fe$_2$As$_2$, the $\kappa_0/T$ may not necessarily manifest systematic change with the increase of the ionic radii of alkali metal.

\section{Summary}

In summary, we have measured the magnetization, resistivity, low-temperature specific heat and thermal conductivity of RbFe$_2$As$_2$ single crystals.
A nodal superconducting gap in RbFe$_2$As$_2$ is strongly suggested by the observation of a significant residual linear term
$\kappa_0/T$ = 0.65 mW K$^{-2}$ cm$^{-1}$ in zero magnetic field. It is concluded that (K,Rb,Cs)Fe$_2$As$_2$ serial superconductors may have a common nodal gap structure, and the field dependence of $\kappa_0/T$ seems to evolve with the impurity level.\\

\begin{center}
{\bf ACKNOWLEDGEMENTS}
\end{center}

This work is supported by the Natural Science Foundation of China,
the Ministry of Science and Technology of China (National Basic
Research Program No: 2012CB821402 and 2015CB921401), Program for
Professor of Special Appointment (Eastern Scholar) at Shanghai
Institutions of Higher Learning. \\

$^*$ E-mail: shiyan$\_$li@fudan.edu.cn

\end{document}